\documentclass[twoside,twocolumn]{article}

\usepackage[hmarginratio=1:1,textwidth=17cm,top=16mm,columnsep=20pt]{geometry}
\usepackage[colorlinks,citecolor=blue]{hyperref}

\title{Science Opportunities and Challenges Associated with SKA Big Data} 
\author{\textsc{Tao AN} \\[1ex] 
\normalsize  Shanghai Astronomical Observatory, 80 Nandan Road, Shanghai 200030, China\\ 
\normalsize \href{antao@shao.ac.cn}{antao@shao.ac.cn} } 
\date{\today}

\begin{document}

\maketitle

The upcoming Square Kilometre Array (SKA) radio telescope will become the largest astronomical observation facility, and is expected to introduce revolutionary changes in major fields of natural sciences. These revolutionary changes help us to answer the fundamental questions related to the origins of the universe, life, cosmic magnetic field, the nature of gravity, and to search for extraterrestrial civilizations \cite{AASKA14}. The unprecedented power of the SKA is characterized by an extremely high sensitivity, wide field of view, ultra-fast survey speed, and ultra-high time, space, frequency resolutions, which ensures that SKA will have a leading position in radio astronomy in future; this will be accompanied by a vast amount of observational data at exabyte (EB) level. The transportation, storage, reading, writing, computation, curation, and archiving of the SKA-level data and the release of SKA science products are posing serious challenges to the field of information communication technology (ICT). The China SKA science team will work together with the information, communication and computer industries to tackle the challenges associated with the SKA big data, which will not only promote major original scientific discoveries, but also apply the obtained technological achievements for stimulating the national economy.

Astronomy is one of the oldest disciplines and has engaged generations of human civilization; China was among the earliest countries to develop astronomy. More than 400 years have passed by since Galileo developed astronomical telescopes, and all the succeeding advancements in astronomy can be attributed to the leap forward in telescope instrumentation and observation techniques. Modern astronomy is inseparable from advanced telescopes. China is in the midst of a strategic opportunity for technological innovation in the new era. Recently, several large-scale astronomical telescopes, such as the Large Sky Area Multi-Object Fiber Spectroscopy Telescope (LAMOST) and the Five-hundred-meter Aperture Spherical radio Telescope (FAST), as well as space-based telescopes such as the DArk Matter Particle Explorer (DAMPE) and the Hard X-ray Modulation Telescope (HXMT), have been developed in China. These facilities are at the threshold of achieving the international top-class level science outputs. Furthermore, China is a prominent participant in the world's largest astronomical scientific project, SKA, which has extensive international cooperation; additionally, SKA will become a flagship radio telescope for natural sciences upon completion and will establish a new milestone in the history of universe exploration \cite{AASKA14,Dewdney}. The existing ground- and space-based telescopes (including space stations) that are capable of X-ray, ultraviolet, optical, infrared and radio cover the full band of observation capabilities. These telescopes are currently being upgraded to achieve major improvements in data quality, which can potentially lead to an exponential growth of the data rates and volumes, and can provide big data for constructing astronomy research. Current astronomical data have reached petabyte (PB) level. With the advancement of the observation technology and with updates of the observation equipment, the data will soon enter the EB level. Astronomical big data will considerably change the manner in which humans explore (including the deployment of machine learning and artificial intelligence for data handling) and understand the nature.


Modern astronomy has managed to produce thought-provoking results, which have  enabled fundamental discoveries in natural science. Exciting and pioneering astronomical discoveries are heavily reliant on the collaborative operation of large-scale scientific research facilities and on the mining and analysis of massive data. The transparency, diversity, and interdisciplinary integration of scientific results have profoundly impacted the manner in which astronomers approach both science and technology. Astronomy has truly entered the era of multi-wavelength and multi-messenger capability. We can now use multiple observing devices for simultaneously detecting the same celestial sources, obtaining their complete electromagnetic spectrum, and using media other than electromagnetic radiation, such as neutrinos and gravitational waves, to study the sources. A recent representative example of such successful coordination was the discovery of the merger of two neutron stars in August 2017. The ground-based Laser Interferometer Gravitational-Wave Observatory (LIGO) initially discovered the space--time ripples produced by the merging of two neutron stars \cite{GW170817}; further, the most powerful space telescopes and ground telescopes were triggered, leading to combined follow-up observations, enabling us to not only detect gravitational waves \cite{LIGO} but also understand their origin (merger scenario), and study their association with gamma ray bursts and giant kilonovae. These advances lend a novel perspective with respect to the power of collaborative astronomical research for pursuing big science.

Observation-based astronomy has long been plagued by data scarcity. In the information age of the 21st century, revolutionary changes have considerably mitigated this scarity. Astronomical observations have now been closely related to big data, with the current scientific research undergoing significant changes. A typical contemporary example is the study of supernovae, which are cosmic explosions that have been studied for a long time; the earliest known event related to supernovae was recorded by ancient Chinese astronomers (the supernova explosion associated with Crab Nebula observed in 1054 C.E.). Supernovae have an important place in modern astrophysics research. The 2011 Nobel Prize in Physics was awarded to three astronomers who presented unambiguous evidence that the universe was accelerating based on the supernovae observations \cite{SNIa}. However, until approximately a decade ago, a supernova was a very difficult event to capture because of the limited technological capabilities at that time. Any serendipitous observation of a supernova  would inevitably attract the attention of global telescopes. Thus considerable number of studies had to rely on numerical simulations and theoretical calculations. Today, large automated optical surveys are discovering more than 1,000 supernovae every year, rendering them less extraordinary while enabling detailed classifications and understanding. These large surveys effectively mine the accumulated data to report additional novel discoveries. In the subsequent 5--10 years, astronomical observations through the next generation of large telescopes will enable the currently rare observation of celestial bodies and events to become more regular. Statistics, information science, and astronomy will work in unison to provide astronomers with data analysis tools. Based on this collection of data, the organization and analysis of big data from cosmic sources, and the physical principles governing their nature and place in the Universe will be extensively explored.
	

Astronomy focuses on pressing issues related to the origins of the universe and its constituents, including galaxies, stars, planets, and life. Breakthroughs and answers to these problems will vastly advance fundamental natural science, promoting the overall progress of science and technology. The SKA radio telescope will enable potential major scientific discoveries; the largest international cooperation project can be observed in the field of astronomy, where the SKA telescope will create new opportunities to unravel cosmic mysteries. The SKA international organization currently comprises 12 member countries, including China, and some associated and observer countries. The construction and operation of the astronomical telescope truly embodies a country's commitment toward the comprehensive development of natural science. The SKA is headquartered in the UK. The SKA low frequency array (SKA-low) includes 1.3 million log-periodic antennas, and will be built in the western Australian desert, whereas the SKA high frequency array (SKA-mid) includes 2,500 dish antennas, which are to be built in eight countries, including and around South Africa. These two locations were selected by astronomers after careful examination and evaluation that spanned for more than a decade. The total collecting area of the telescope is up to one square kilometer, and a wide and finely sampled frequency range of 50 MHz--20 GHz can be covered.
	
The SKA will play a leading role among the next generation of radio astronomy observation facilities, having novel capabilities including ultra-high sensitivity (milli-Kelvin), large field of view (thousands of square degrees), ultra-fast survey speed, ultra-high frequency resolution (kilo-Hz), ultra-high time resolution (nanoseconds), and ultra-high spatial resolution (sub-arcseconds).  The construction of the SKA is mainly divided into two phases. The first phase (SKA1) will cover approximately 10\% of the full scale with construction expected to commence in 2020. The sensitivity of the SKA1 in the centimeter band will be $\sim$50 times higher, and its sky survey speed will be $\sim$10,000 times higher than that of the existing largest radio telescope array \cite{AASKA14}. Further, the scientific objectives and overall engineering schemes of the SKA1 have been clearly defined. The second phase (SKA2) will cover the remaining 90\% based on a detailed plan which is currently under discussion.

The aforementioned technical advantages will inevitably enable the SKA to produce unprecedented data volumes. The data rate from the SKA1-low stations to the correlator will be $\sim$10 terabits per second (Tb/s) and will require in total $\sim$1 peta operations per second, producing the largest data stream to date. In comparison, the SKA2 is expected to generate at least 10 times more real-time data streams than those generated by the SKA1. Even after pre-processing and post-correlation, the amount of pre-calibrated data will be considerably large. The input data rate into the scientific data processors (SDPs) in Australia and South Africa will become 4 giga bits per second. SKA's ultra-large-scale data flow must be processed in real-time to ensure the steady operation of the data processing pipeline without any blockages. The use of a real-time mode and multiple concurrent tasks makes the data stream processing pipeline a typical feature of SKA, providing extensive applicability to generic large-scale scientific computing.
	
As the largest radio telescope in history, SKA's mission will not only include making world-class scientific discoveries but also producing the world's largest data archive. Thus, scientists have to completely understand the upcoming huge challenges that will be posed by SKA data processing. Owing to the extremely large project size and intricate components, various pathfinders and precursor facilities, including Chinese 21CMA, have been built in several countries \cite{NewScientist}  to reduce or overcome the technological shortcomings;  each project is equivalent to approximately 1\% of the overall scale of the SKA. Based on the operation of these pathfinder telescopes, relevant science and technology research is being conducted. These pioneering facilities have played a fundamental role in understanding SKA's scientific goals, establishing and gradually improving the sky model, developing and testing the data processing software, and cultivating the much-needed talent teams.  However, a major shortcoming is that the current data volumes from these pathfinders are considerably less than the SKA1 scale; consequently, they have long distances to span for  providing real verification.


When compared with traditional telescopes, SKA is the closest to being involving (in addition to the expected science objectives) an intricate integration that considers the latest achievements of contemporary information and computing technology; however, this raises new problems, particularly related to big data. For example, SKA-low is designed to detect weak signals from the Cosmic Dawn. The low-frequency arrays will produce the world's largest data stream at ten petabits per second, which far exceeds the global internet traffic. According to the SKA data flow design, the amount of scientific data that will be transported to the regional data centers for performing in-depth analysis during the first phase of construction will reach 600 PB /yr. By completely operating the telescope, the amount of data could exponentially  increase. By the era of SKA2, the data generated from the SKA telescopes will extend to become more than 100 times that of the SKA pathfinders, reaching the exabyte (EB) scale. SKA's two most important scientific objectives, which are exploring the epoch of re-ionization and cosmic dawn and the discovery of gravitational waves with pulsar timing arrays, require an accumulation of raw data; considering the preservation of raw data for a certain period of time (e.g., 6 months), the SKA observatory data storage requirements will increase by at least a certain order of magnitude.
	
Next, let us consider the SKA precursor project, MWA \cite{MWA}, as an example. After 4 years of operation, MWA has accumulated 24 PB of scientific archive data. One of the key science projects is the Galactic and Extragalactic All-sky MWA (GLEAM) survey. The first phase of the survey has already found more than 300,000 galaxies  \cite{GLEAM}, and the archived data volume has reached more than 1 PB. Based on the second phase of the survey (which is already underway), the sensitivity will increases 4-fold, and the data volume is expected to become as high as 6.5 PB. Because MWA only accounts for 1\% of the SKA-low scale, the amount of SKA data can be approximately estimated to be at the EB level. According to preliminary estimates, the scientific data processor of the SKA1 stage requires at least 260 PFlops, which is equivalent to eight times that of China's supercomputer Tianhe-2, about approximately three times that of the Sunway TaihuLight. SKA's huge computing requirements will inevitably impact the existing computer architecture and the scientific computing algorithms. The developed solutions that intend to solve the data processing requirements of the SKA will help to promote advancements in related industries and considerably influence the general technological developments.

The SKA, which is currently supported by 12 full-member countries, including China, has attracted several hundred scientists and engineers from more than 20 countries for conducting collaborative studies. Chinese engineers joined six of the ten work packages of the SKA \cite{Peng2012,Peng2017}, including the reflector dish for the SKA-mid \cite{Du}, low-frequency aperture array \cite{WuMQ}, mid-frequency aperture array, signal and data transport, science data processor, and wideband single pixel feeds \cite{WuY}.  China plays a leading role in the development of the SKA dish, and the first SKA-mid prototype telescope of 15-m diameter was completely assembled in Shijiazhuang, China on February 7, 2018. Chinese scientists actively participate in 13 associated science working groups and focus groups. China has proposed a customized science strategy by focusing on the exploration of the cosmic dawn and strong-field tests of gravity using pulsars and black holes. In addition, development of the science data processing capabilities in the context of the SKA big data is of considerable importance for realizing the aforementioned objectives. China is currently preparing a prototype of the China SKA data center and is working toward an understanding and finding solutions to  the problems associated with SKA dataflow. This pioneering work will pave the way for the community to quickly produce exciting scientific discoveries from SKA when the SKA1 becomes completely operational by 2028.

In practice, SKA will considerably promote and influence several fields related to astronomy, computer science, informatics, and electronics. High-speed digital sampling and real-time digital signal processing at the Tbps level poses new challenges to the electronics industry. Long-duration work in harsh desert environments and the long-distance transmission of the frequency synchronization are major technical challenges that have to be urgently solved by the aperture array. Large-volume data transferring over long distances (thousands of kilometers) and at high speeds is an area of active ongoing research with respect to the development of intercontinental network infrastructure, topology, and protocols. The software on the  transmission side has stringent requirements for ultra-high-speed data streaming and is highly complex. Increasing the number of interconnected network ports of nodes and the total bandwidth between nodes can offer effective solutions to address this problem; this increase will also promote the layout and construction of the domestic 100 Gbps-level, and even Tbps-level fundamental networks.

SKA data processing, characterized by data-intensive scientific computing, places a higher demand on the electronics, computer, and signal processing industries. One of the biggest challenges of SKA science data processing is the Input/Output (I/O) problem \cite{DALiuGE}. The transmission I/O bandwidth has been observed to be one of the main bottlenecks in the data flow system. Even supercomputers, such as Tianhe-2, will be insufficient for processing the SKA big data, because they are limited by their I/O bandwidth. Thus  developing a new architecture system that adapts to data-intensive scientific computing is important. As mentioned above, SKA's high-speed and massive input data must reduce the pressure of subsequent processes by appealing to real-time processing. The real-time processing of large data volumes demands for the upgradation of the hardware and software systems; the overall system architecture design, and integrated installations. The execution framework of the supercomputing center, supporting software algorithms, data center health monitoring, cabinet cooling, and total control management, will all face big challenges of the tremendous dataflow. In the case of a construction funding cap, considering the predetermined computing power and real-time capabilities is important. However,  long-term operation requires low power consumption, necessitating a new type of computer architecture. In addition, the storage, archiving, retrieval, and calculation involved with massive data places stringent requirements on the complete ecological chain of supercomputers, which differs from the conventional CPU chips that have been deployed in large supercomputers. Although the computer industry has developed deep-learning processor chips for artificial intelligence, current mainstream operating systems, storage systems, and other software ecosystems still lag behind the demands of big data applications. Thus, the SKA project has created a strong drive for the related ICT industries.

In addition to hardware issues, the current status of development of astronomy applications is far from satifying the requirements of SKA. Key algorithms for SKA scientific data processing are based on resources that include shared file systems. Traditional multi-core computer systems often compete for resources during multi-tasking, multi-concurrency, and multi-thread parallel execution. If the data flow execution framework cannot effectively resolve resource scheduling and allocation, it will  halt the data processing pipeline in the worst case. This issue has been recognized as a common problem affecting the SKA precursor telescope data processing center. To address this issue, the Australian ICRAR institute and the Shanghai Astronomical Observatory of the Chinese Academy of Sciences jointly developed a novel data flow execution framework that can be referred to as the Data Activated Stream ({\it Liu}) Graph Engine (DALiuGE) for the SKA project \cite{DALiuGE}, which adopts an advanced ``data-driven" concept. DALiuGE has been confirmed to be more suitable for SKA scientific computing than for the traditional HPC ``computation-driven" design. In addition, the existing computational efficiency of SKA software programs is less than 10\%, which is far lower than what was originally planned. Consequently,  the theoretical peak performance of 260 PFlops is far from satifying the actual requirements of scientific data processing. The simple solution of increasing the supercomputing cabinets is not practical. Instead, a more feasible approach is to improve the software efficiency. Even increasing the computing efficiency from 10\% to 20\% will entail the savings and significantly reduce the operation expenses. Astronomers need to work closely with computer experts to optimize the algorithms for ensuring speed, robustness, and scalability. Hence, cultivating a diverse talent pool that understands both astronomy and computational algorithms is essential. Another practical problem involves astronomical data processing software, which requires urgent updates to satisfy the future requirements. Majority of the current astronomy software packages have been developed in the 1970s and 1980s. Modern instrumentation and telescopes demand accounting for high-speed, real-time, and parallel processing of big data. Astronomers have begun to use increasingly advanced modular and parallel development support languages. An alternative version of the Astronomical Image Processing System (AIPS) \cite{AIPS} developed by the National Radio Astronomy Observatory of the US, which is called the Common Astronomy Software Applications (CASA) \cite{CASA} (developed in C++), will characterize the transition to the next generation of mainstream radio astronomy software. Further, programs involving machine learning and artificial intelligence are expected to be prioritized. The development of astronomy data processing software, such as astronomical research, has been transformed from a development involving  scattered individuals to one with global cooperative operation. For example, the LIGO team that discovered the gravitational waves is a large team containing  more than 1,000 scientists. There are almost 100 personnel (from around the world) contributing to the algorithms that are extensively used in the CASA core library. Thus, large teams possessing a diverse range of talents will be essential to tackle apparent roadblocks and ensure successful operation for enabling novel discoveries.

Apart from exploring the science frontiers, science communication is also very important for providing information at all levels (e.g., users of the facility, interested industries, and governments, and the public) and for attracting relevant talent.  In the future, SKA's astronomical big data will not only serve astronomers but will also provide an interface to the public. In conjunction with conducting novel scientific research, we must promote important research results, popularize science and technological knowledge,  and promote scientific spirit which involves a basic awareness of the scientific principles and their acceptance by the general public. Through the virtual observatory and the public ``cloud" the SKA Regional Data Centers will allow convenient access to frontier science, enabling its popularity among general users.
	
SKA will dominate and influence  developments in radio astronomy for the next 50 years and will influence vigorous advances in the low-frequency radio domain. It will catalyze major scientific breakthroughs, including the study of observational cosmology. SKA is the largest project globally involving international cooperation and has created an important opportunity for China's radio astronomy teams to achieve a leading future role.

According to the design of the SKA, in-depth processing and analysis of SKA data is expected to be undertaken by several regional data centers distributed across several continents. Major member countries, including China, have initiated a concept study and prototyping of the SKA Regional Data Center.
With such high expectations, preliminary technical research and prototyping work have already begun. Owing to the complexity and large data volumes as well as rates in SKA data processing, frequent movement of this large-scale data is considered to be impractical. Consequently, centralized data processing in major regional data centers is inevitable. Building the China SKA Regional Data Center is not only an indispensable part of the international SKA project, but is also an important guarantee to support the Chinese scientists for effectively using the SKA data to obtain novel scientific returns. Because the scientific users of the SKA are located at  diverse locations around the world, distributed computing and storage along with cloud computing and archiving become important considerations. The regional center network, comprising multiple science and data sub-centers, can satify the diverse requirements of the SKA user community.

Currently, an end-to-end direct optical fiber connection has been established between the data centers at the Shanghai Astronomical Observatory and the SKA data center in Perth, Australia. The maximum data transmission rate is currently 3.2 Gbps, which is the highest known astronomical data flow rate in China; this rate has provided a useful practical model for future operations in the context of the SKA Regional Data Center network.\footnote{At the time of writing this article, the total network bandwidth between Shanghai (China) and Perth (Australia) has increased to 10 Gbps.} Multiple data attributes that characterize the diverse range of SKA scientific applications render parallelized multi-data streaming as an inevitable consequence (as mentioned in Section 3), which requires concentrated attention while  developing a future international network of SKA regional data centers.

To keep pace with the frontier developments at the international level, we must not only independently approach the oncoming problems but also cooperate with leading international institutes and world-class research teams to improve and efficiently employ the existing capabilities and to plan further upscaling. China currently lacks a pool of skilled scientists processing data and must completely recognize the long-term beneficial nature of talent training. China's SKA science team should seize the valuable time window before SKA1's first batch of data is released around 2024, focus on relevant scientific research, use the SKA pathfinder/precursor telescopes to discover novel scientific directions, and master the data processing technology.  In addition to astronomical research and data processing talents, management-oriented experts with science and technology backgrounds are highly relevant for the consolidation and presentation of key science and technology outputs  in large-scale international collaborations.

In response to the challenges posed by SKA big data, the China SKA science team should firmly develop international cooperation and accelerate the localization of key technologies to ensure mutual growth and independent verification. The development of an SKA regional science and data center in China  intends to achieve breakthroughs in key technologies, such as Tbps-level high-speed scientific research networks, signal and data transferring, EB-level high-performance computing and storage, and supporting astronomical software for diverse scientific applications. This will ensure that the incoming data streams can be used to quickly achieve significant scientific results that will propel further rapid scientific advancements upon the arrival of the SKA era.

In conclusion, because we all share a common sky, participating in the SKA global collaboration is a giant leap toward answering the fundamental questions posed by astronomy; further, this participation also contributes to the solution of the cosmic mysteries at the frontiers of science. Thus, the SKA global collaboration is an important measure/task for ``building a community of common destiny for mankind" as proposed by the Chinese government.

{\bf Acknowledgements} This work is supported by the National Key R\&D Program of China (Grant No. 2018YFA0404600) and the International Cooperation Bureau of the Chinese Academy of Sciences (Grant No. 114231KYSB20170003). TA thanks Prashanth Mohan for his help to prepare the manuscript.

\end{document}